\documentclass[twocolumn,reprint,aps,superscriptaddress,prl]{revtex4-2}
\usepackage[utf8]{inputenc}
\usepackage{color}
\usepackage{amsmath}
\usepackage{amssymb}
\usepackage{amsbsy}
\usepackage{bbold}
\usepackage{graphicx}
\usepackage{xfrac}
\usepackage{physics}
\usepackage{textcomp}
\usepackage[]{todonotes}
\usepackage[unicode=true,
 bookmarks=false,
 breaklinks=false,pdfborder={0 0 1},backref=false,colorlinks=true]
 {hyperref}
\hypersetup{
 linkcolor=blue,citecolor=blue,urlcolor=blue,linkcolor=blue,citecolor=blue,urlcolor=blue}

\makeatletter

\usepackage{braket}

\definecolor{mygreen}{rgb}{0,0.5,0}
\definecolor{myblue}{rgb}{0,0,0.75}
\definecolor{mymagenta}{cmyk}{0,1,0,0.12}

\makeatother

\begin{document}

\title{Second-harmonic generation via double topological valley-Hall kink modes in all-dielectric photonic crystals}

\author{Zhihao Lan}
\affiliation{Department of Electronic and Electrical Engineering, University College London, Torrington Place, London, WC1E 7JE, United Kingdom}
\author{Jian Wei You}
\affiliation{Department of Electronic and Electrical Engineering, University College London, Torrington Place, London, WC1E 7JE, United Kingdom}
\author{Qun Ren}
\affiliation{Department of Electronic and Electrical Engineering, University College London,
Torrington Place, London, WC1E 7JE, United Kingdom} \affiliation{School of Electrical and
Information Engineering, Tianjin University, 300072, Tianjin, China}
\author{Wei E. I. Sha}
\affiliation{Department of Electronic and Electrical Engineering, University College London, Torrington Place, London, WC1E 7JE, United Kingdom}
\affiliation{Key Laboratory of Micro-Nano Electronic Devices and Smart Systems of Zhejiang Province, College of Information Science and Electronic Engineering, Zhejiang University, Hangzhou 310027, China}
\author{Nicolae C. Panoiu}
\affiliation{Department of Electronic and Electrical Engineering, University College London, Torrington Place, London, WC1E 7JE, United Kingdom}

\begin{abstract}
Nonlinear topological photonics, which explores topics common to the fields of topological phases
and nonlinear optics, is expected to open up a new paradigm in topological photonics. Here, we
demonstrate second-harmonic generation (SHG) via nonlinear interaction of double topological
valley-Hall kink modes in all-dielectric photonic crystals (PhCs). We first show that two
topological frequency bandgaps can be created around a pair of frequencies, $\omega_0$ and
$2\omega_0$, by gapping out the corresponding Dirac points in two-dimensional honeycomb PhCs.
Valley-Hall kink modes along a kink-type domain wall interface between two PhCs placed together in
a mirror-symmetric manner are generated within the two frequency bandgaps. Importantly, through
full-wave simulations and mode dispersion analysis, we demonstrate that tunable, bi-directional
phase-matched SHG via nonlinear interaction of the valley-Hall kink modes inside the two bandgaps
can be achieved. In particular, by using Stokes parameters associated to the magnetic part of the
valley-Hall kink modes, we introduce a new concept, SHG directional dichroism, which is employed to
characterize optical probes for sensing chiral molecules. Our work opens up new avenues towards
topologically protected nonlinear frequency mixing and active photonic devices implemented in
all-dielectric material platforms.
\end{abstract}
\maketitle

\textit{Introduction.---} Recent advances in topological photonics \cite{review_NP14, review_NP17,
review_PQE17, review_AOM17, review_OE18, review_JAP19, review_RMP19} have led to new ways to
control light in a robust manner using photonic states that are protected by the topological
properties of the systems. Earlier works \cite{QHE_PRL08_Haldane, QHE_PRL08_Wang,
QHE_Nature09_Wang, QHE_PRL11_Poo} in this field have focused on the realization of photonic
analogue of quantum Hall states in two-dimensional (2D) photonic crystals (PhCs) containing
magneto-optical materials, where the time-reversal symmetry is broken by external magnetic fields.
As magneto-optical effects are generally weak at optical frequencies, intense research efforts were
devoted to topological photonic systems without magneto-optical materials and concepts such as
Floquet topological phases and synthetic magnetic fields have been demonstrated in helical
waveguide arrays \cite{Floquet_Nature13_Rechtsman} and coupled ring resonators
\cite{Loop_NP13_Hafezi}, respectively. Furthermore, photonic systems emulating quantum spin Hall
\cite{QSH_NM13_Khanikaev, QSH_PRL15_Ma, QSH_PNAS16_He, QSH_15PRL_Wu} and quantum valley Hall
\cite{QVH_NJP16_Ma, QVH_NC19_He,QVH_NN19_Shalaev,QVH_LPR19_Ma} effects, which preserve the
time-reversal symmetry of the system, have also been proposed. We note that for all systems
discussed above, the topological photonic properties can be understood within the single-particle
framework.

On the other hand, interacting topological phases provide an exciting topic in condensed matter
physics \cite{review_RPP18_Rachel} and in the context of photonics, the existence of nonlinearity
in many optical materials \cite{book_Boyd08} provides a unique platform to study interaction
effects in topological physics, which is expected to greatly expand our understanding of
topological photonic systems \cite{review_NTP}. Indeed, lattice edge solitons
\cite{solition_PRL13_Lumer, solition_PRL16_Leykam, solition_Mukherjee}, nonlinear control
\cite{control_PRL19} and imaging \cite{imaging_PRL19} of photonic topological edge states,
traveling-wave amplifiers \cite{amplifier_PRX16}, topological insulator lasers
\cite{TIlaser_Science18T, TIlaser_Science18E}, topological sources of quantum light
\cite{quantum_Nature18}, and the potential to enhance harmonic generation \cite{harmonic_OE18,
harmonic_NC19, harmonic_NN19} have been demonstrated. Despite these advances in understanding the
nonlinear effects in topological photonic systems, achieving nonlinear frequency mixing -- one of
the fundamental nonlinear optical processes -- via phase matching topological edge states is still
largely unexplored. Recently, we have studied four-wave mixing of topological edge plasmons in
graphene metasurfaces \cite{SciAdv20_You} and second- and third-harmonic generation (SHG, THG) in
topological PhCs \cite{PRB20_Lan}, using one-way edge modes similar to quantum Hall states with
external magnetic field. Nevertheless, whether the above goal can be realized in time-reversal
symmetry-preserving topological photonic systems without exploiting magneto-optical effects is
still an open question, which we address here.

In this Letter, we demonstrate SHG in all-dielectric PhCs, through nonlinear interaction of
topological edge modes within two different frequency bandgaps around $\omega_0$ and $2\omega_0$.
Our implementation is based on the photonic quantum valley Hall effect and its associated
valley-Hall kink modes \cite{QVH_NJP16_Ma, QVH_NC19_He,QVH_NN19_Shalaev,QVH_LPR19_Ma}. A key
novelty of our work lies in the design of two topological valley gaps hosting double valley-Hall
kink modes that can be phase matched to achieve SHG. Importantly, unlike the case of one-way edge
modes \cite{SciAdv20_You, PRB20_Lan}, in the current system one could launch the fundamental wave
along either directions of the topological interface, thanks to the topology of the valley-Hall
kink modes, with second harmonic waves being generated both in the forward and backward direction
due to the time-reversal symmetry of the system. Moreover, a unique feature of our system relevant
to many applications is that the amplitudes of these two wave components can be readily tuned by
varying the frequency or source location and chirality, a functionality that non-topological
nonlinear optics cannot provide.
\begin{figure}
\includegraphics[width=\columnwidth]{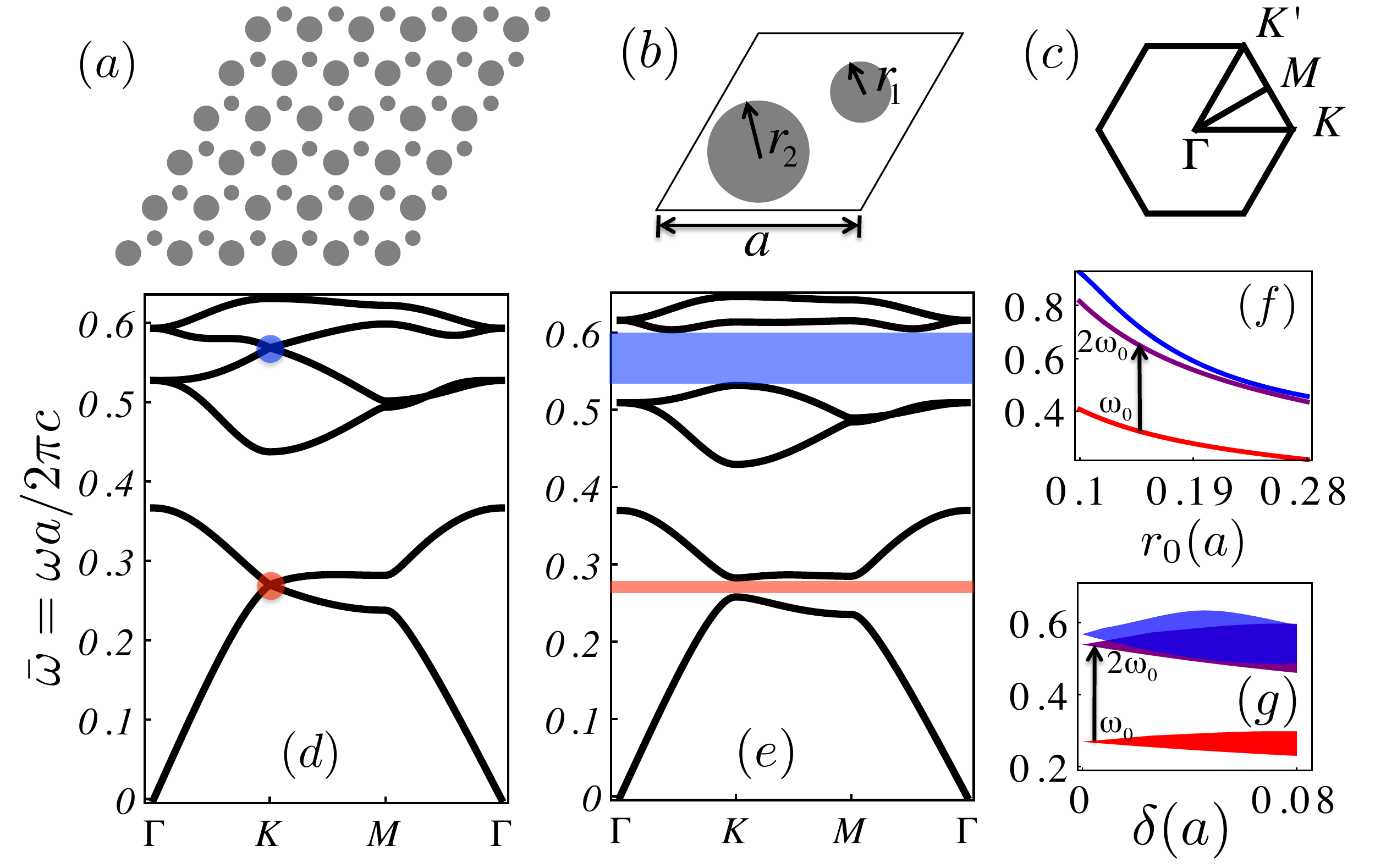}
\caption{\textit{Emergence of two valley gaps for SHG.} (a) Schematic of the system. (b) Unit cell
of the PhC, containing two cylinders of nonlinear material with radius $r_1$, $r_2$, dielectric
constant $\epsilon=12$ and $\chi^{(2)}=10^{-21}$CV$^{-2}$. (c) First Brillouin zone of the PhC. (d)
The existence of double Dirac points (marked by red and blue dots) of the PhC at
$r_{1}=r_{2}=0.2a$. (e) The same as in (d), but calculated for $r_1=0.18a$ and $r_2=0.22a$. (f) The
frequencies of the two Dirac points in (d) vs. $r_0$ when $r_{1}=r_{2}=r_0$. (g) The width of the
two valley gaps in (e) vs. $\delta$, defined as $r_1=r_0-\delta$ and $r_2=r_0+\delta$ with
$r_0=0.2a$. \label{fig:fig1}}
\end{figure}

\textit{The system.---} To demonstrate the main ideas, we consider a 2D
honeycomb PhC made of dielectric cylinders with radius $r_{1}$ and $r_{2}$, see
Fig.~\ref{fig:fig1}(a), whose unit cell and first Brillouin zone are shown in
Figs.~\ref{fig:fig1}(b) and \ref{fig:fig1}(c), respectively. The cylinders are made of nonlinear
material with dielectric constant $\epsilon$ and second-order nonlinear susceptibility
$\chi^{(2)}$. In the following, we use normalized frequency and momentum, $\overline{\omega}=\omega
a/2\pi c$ and $\overline{k}=k a/\pi$, with $c$ the speed of light and $a$ the lattice constant. The
transverse magnetic (TM) modes of the honeycomb PhC possess Dirac points between the first and
second bands \cite{HoneyValley_PRB17dong, HoneyValley_SciRe18Hang}, a feature exploited for
topological valley transport. However, Dirac points at higher bands have been less studied. We show
in Fig.~\ref{fig:fig1}(d) the first six TM bands, determined using
BandSOLVE\texttrademark~\cite{RSoft}, when $r_{1}=r_{2}\equiv r_0$, from which one can see the
existence of double Dirac points, whose frequencies as a function of $r_0$ are shown in
Fig.~\ref{fig:fig1}(f).
\begin{figure}
\includegraphics[width=\columnwidth]{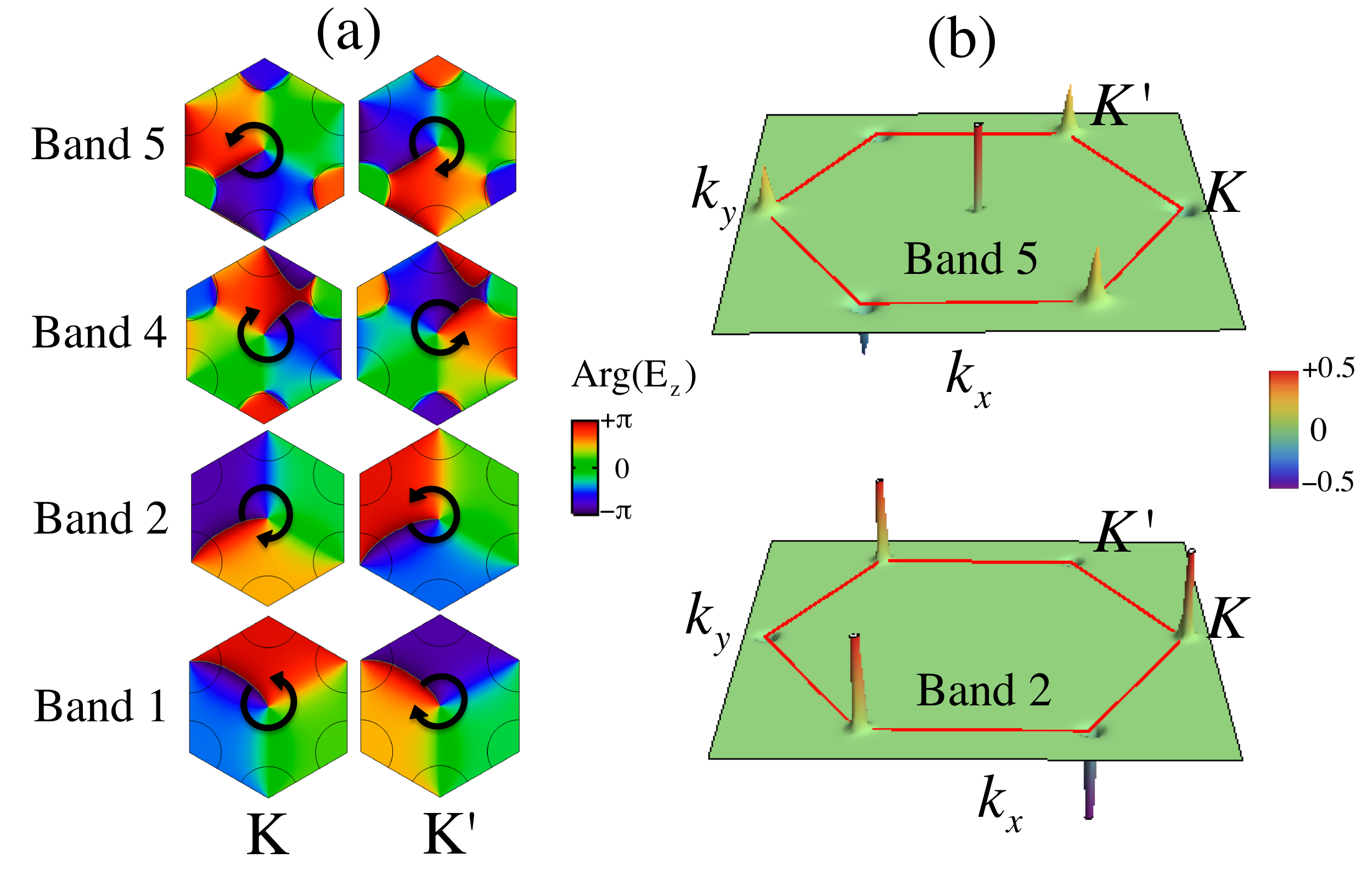}
\caption{\textit{Topological properties of the two valley gaps.} (a) Phase winding behaviors of
$E_z$ at $K$ and $K'$ for the four bands around the Dirac points in Fig.~\ref{fig:fig1}(d). (b)
Berry curvature distributions of band 2 and 5 (Berry curvatures of band
1 and 4 around $K, K'$ show opposite behaviors to band 2 and 5, thus not shown). The peak
of Berry curvature at the $\Gamma$ point of band 5 is due to the band degeneracy between band 5 and
6 at $\Gamma$.  In the simulations, a small cylinder difference is used to gap out the Dirac points
and we have checked the integral of the Berry curvature around $K$ and $K'$ gives
$\pm\pi$. \label{fig:fig2}}
\end{figure}

The Dirac points could be gapped out by using a unit cell containing cylinders with
different radius. We show in Fig.~\ref{fig:fig1}(e) the band structure of the PhC with $r_1=0.18a$
and $r_2=0.22a$, from which one can see the gapping out of the Dirac points by forming valley gaps.
The effect of the inversion symmetry breaking can be quantified by $\vert r_2-r_1\vert$. We further
show in Fig.~\ref{fig:fig1}(g) the width of the two valley gaps when varying the radius difference.
As can be seen, the second-harmonic gap (purple) with respect to the first valley gap (red)
overlaps significantly with the second valley gap (blue), thus the two valley gaps (red and blue)
can be used for SHG.
\begin{figure}
\includegraphics[width=\columnwidth]{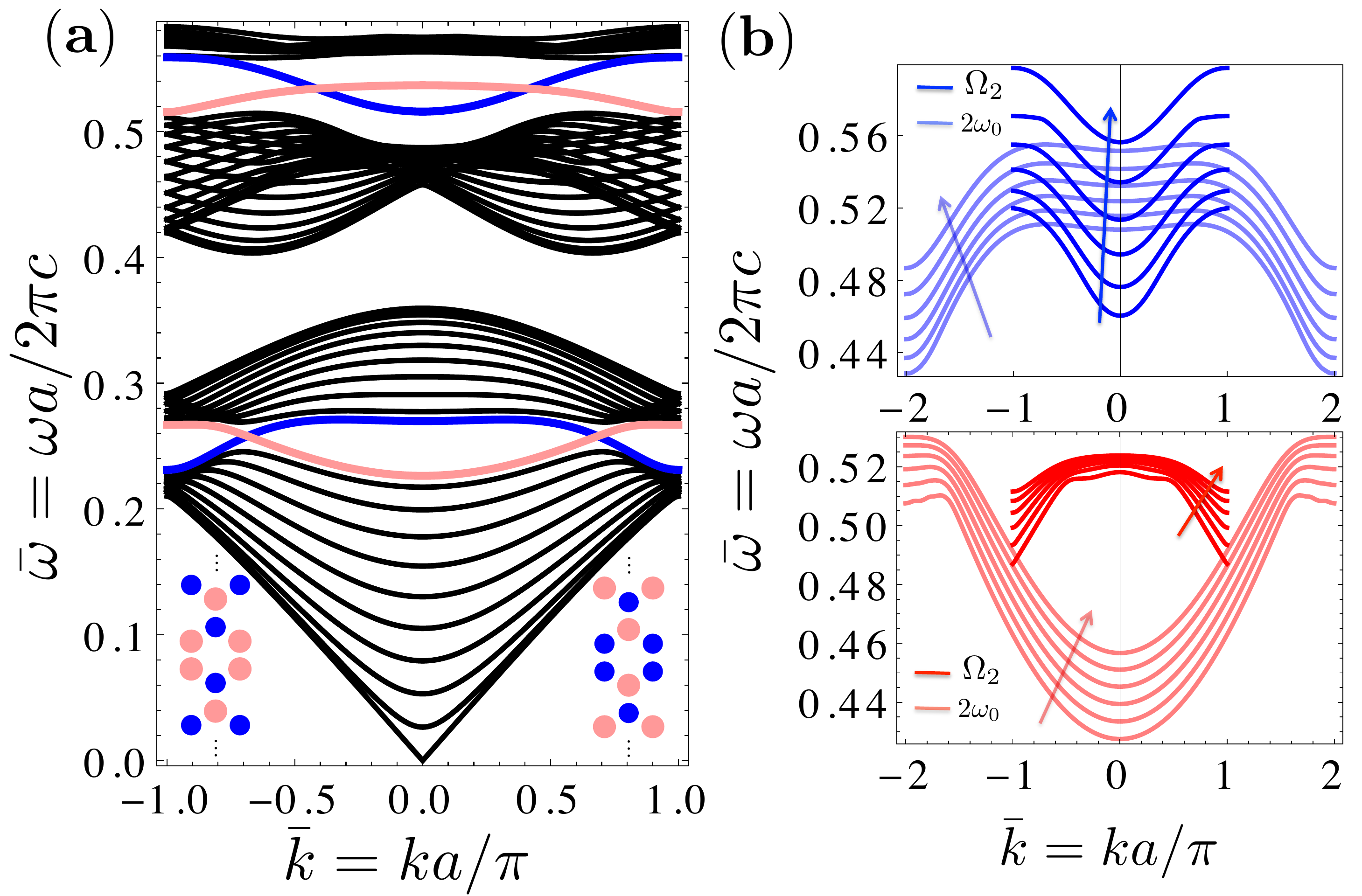}
\caption{\textit{Emergence of valley-Hall kink modes in the two valley gaps for SHG.} (a) Band
structure of a kink-type domain wall interface separating two PhCs (with $r_1=0.2a, r_2=0.23a$)
that are mirror symmetric to each other (see the inserts). Note the red (light gray) and blue (dark gray) kink modes
correspond to the interface with bigger cylinders (red, left insert) and interface with smaller
cylinders (blue, right insert), respectively. (b) Tuning the kink mode dispersion to achieve phase matching for SHG, where the
fundamental modes $\omega_0$ are shown in the second-harmonic gap by applying the transformation
$(\omega_0, k_f)\mapsto(2\omega_0, 2k_f)$ and $\Omega_2$ refers to the second harmonic modes. Along
the arrow direction, $r_1$ varies from $0.23a$ to $0.18a$ with step of $0.01a$, whereas
$r_2=0.24a$. \label{fig:fig3}}
\end{figure}

\textit{Topological properties of the valley gaps.---}  To demonstrate the topological nature of the two valley gaps,
we present in Fig.~\ref{fig:fig2} the phase winding behavior of
$E_z$, and the Berry curvature distribution around the two valleys $K$ and $K'$
\cite{Chern_ZhaoOE20, Chern_LuFO20, Chern_PazAQT20}. Figure~\ref{fig:fig2}(a) shows  that the phase
winding behavior of $E_z$ for the first valley gap between bands 1 and 2 is opposite to that of the
second valley gap between bands 4 and 5. Moreover, they are opposite to each other at the $K$ and
$K'$ valleys for all four bands.

The band topology is characterized by the Berry curvature
$\mathcal{F}(\mathbf{k})=\nabla_{\mathbf{k}}\times \mathcal{A}_n(\mathbf{k})$, where
$\mathcal{A}_n(\mathbf{k})=\expval{i\nabla_{\mathbf{k}}}{u_{n\mathbf{k}}}$ is the Berry connection
with $\ket{u_{n\mathbf{k}}}$ being the eigenmode of the $n$-th band. The integral of the Berry
curvature over the first Brillouin zone modulo 2$\pi$ gives the Chern number and this integral is
zero for systems with time-reversal symmetry \cite{QHE_PRL08_Wang}. However, the Berry curvature
can have nontrivial local distributions around $K$ and $K'$, which can be used to define the valley
Chern number ($\pm1/2$). From Fig.~\ref{fig:fig2}(b), one can see that the Berry curvature
distributions have opposite sign at $K$ and $K'$ and are opposite for the first and second valley
gaps, too. As we will show later on, these features have important implications on the kink
 modes induced within the two gaps around $K$ and $K'$.

\textit{Emergence of chiral valley-Hall kink modes within the two valley gaps.---} According to the
bulk-edge correspondence principle, for an interface separating two bulk systems, if the difference
of the topological invariants of the bulk systems across the interface is nontrivial, interface
states will emerge inside the bulk bandgap. For our case, if one PhC (I) is inversion symmetric to
the other (II), the two valleys will be transformed to each other, i.e.,
$K_{I}/K^{\prime}_{I}=K^{\prime}_{II}/K_{II}$. As the valley Chern number at $K$, $K^{\prime}$ is
$\pm 1/2$, see Fig.~\ref{fig:fig2}(b), $C_{K/K^{\prime}}^{I}=-C_{K/K^{\prime}}^{II}=\pm 1/2$ and
consequently $C_{K/K^{\prime}}^{I}-C_{K/K^{\prime}}^{II}=\pm 1$. This means that at one valley
there exists an interface mode with positive group-velocity (GV) and another one with negative GV
exists at the other valley. Importantly, as the two valley gaps have opposite topologies, see
Fig.~\ref{fig:fig2}, the interface modes within the two valley gaps at the same valley have
opposite GV, a unique feature for achieving phase matching. The band structure of a kink-type
domain wall interface is calculated and presented in Fig.~\ref{fig:fig3}(a). One can see that, the kink modes
within the two valley gaps indeed have
opposite GVs (slopes) at a specific valley (for further topological properties of the valley modes,
see \cite{supp_mat}).
\begin{figure*}
\includegraphics[width=\textwidth]{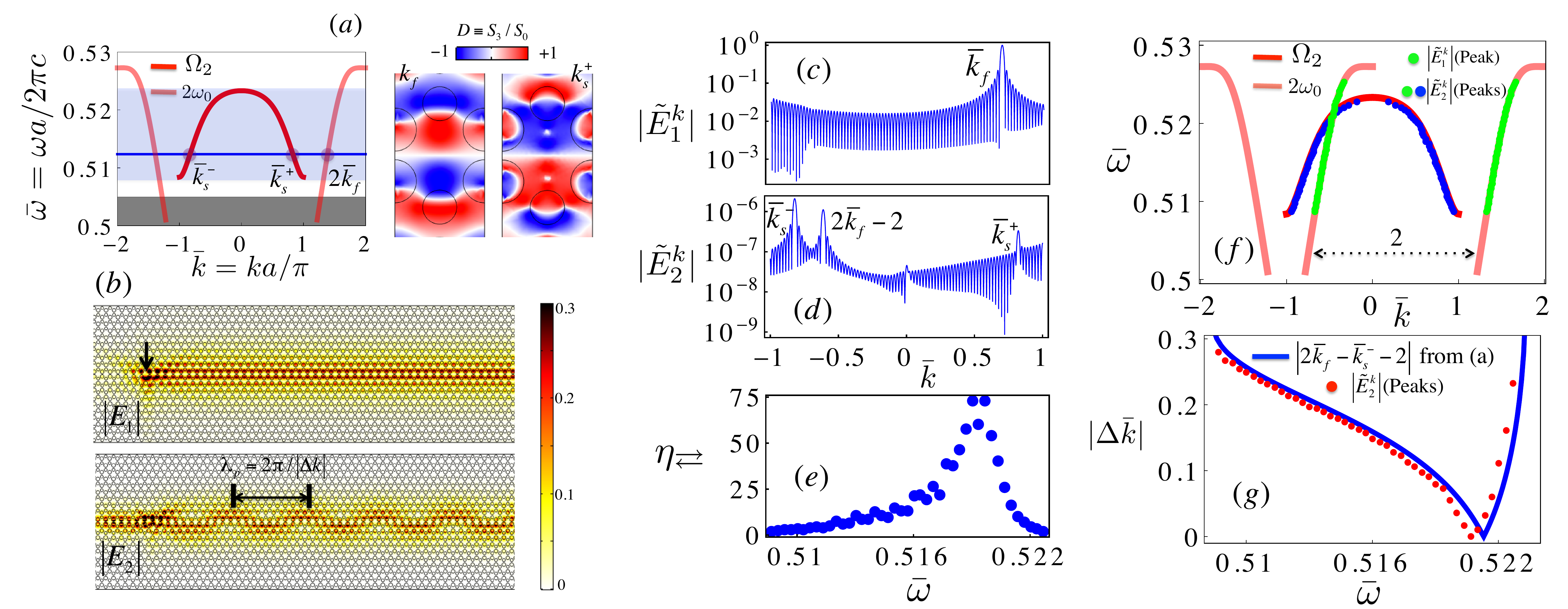}
\caption{\textit{SHG via double valley-Hall kink states.} (a) Dispersion curves (left) used for the
SHG (from bottom panel of Fig.~\ref{fig:fig3}(d) at $r_1=0.19a, r_2=0.24a$), where the shaded grey
area is occupied by the bulk modes whereas the light blue indicates the frequency matching window
and chirality maps (right) of the valley modes at $k_f$ and $k_s^+$. (b) Simulated field
intensities of the fundamental ($E_1$) and second-harmonic ($E_2$) waves at the frequency marked by
the blue line in (a). The fundamental wave is excited unidirectionally via a chiral source located
at the extremum of the $k_f$ chirality map, which is realized by six dipoles with phase winding in
the simulations. (c), (d) Fourier transform of the fields $E_1$ and $E_2$ in (b) (both are
normalized by the maximum of $|\tilde{E}_1^k|$ ). (e) The forward-to-backward ratio
($\eta_{\rightleftarrows}$) of the generated second-harmonic waves vs. frequency determined from
the amplitudes of the peaks at $\bar{k}_s^-$ and $\bar{k}_s^+$ in (d). (f) Extracted dispersion
curves from the peaks of $|\tilde{E}_1^k|$ and $|\tilde{E}_2^k|$ (green for the fundamental wave
and blue for the harmonic wave) in (c) and (d) compared to the dispersion curves of (a). (g)
Extracted $|\Delta \bar{k}|$ from the peaks of $|\tilde{E}_2^k|$ in (d) compared to that obtained
from the dispersion curves of (a). \label{fig:fig4}}
\end{figure*}

To tailor the kink modes for efficient SHG, we show in Fig.~\ref{fig:fig3}(b) the evolution of the kink mode dispersion
curves when changing the radius difference of $r_1$ and $r_2$. When increasing $\vert
r_{1}-r_{2}\vert$, as indicated by the arrow direction, the widths of the two valley gaps increase
and the dispersion curves move towards higher frequencies. As we require the bandgaps to be
relatively large, one could expect to achieve efficient SHG at large radius difference. From
Fig.~\ref{fig:fig3}(b), one can see that when $\vert r_{1}-r_{2}\vert$ is large, the dispersion
curves of the fundamental and second harmonic kink modes corresponding to the smaller cylinder
interface (top panel) move away from each other, thus phase matching can not be realized
effectively in this case. Conversely, the dispersion curves corresponding to the larger cylinder
interface (bottom panel) are well overlapped for large radius difference, which is ideal for
fulfilling phase matching requirements. Importantly, Fig.~\ref{fig:fig3}(b) suggests that near the
$\Gamma$-point it is possible to achieve phase matching when both the fundamental and
second-harmonic waves are in the slow-light regime $(v_{g}\ll c)$, which results in greatly
enhanced SHG.

\textit{SHG via double valley-Hall kink modes.---}  For clear illustration of the SHG, we present
both the fundamental and second-harmonic kink modes in the second valley gap, see
Fig.~\ref{fig:fig4}(a). Due to the time-reversal symmetry of our system, at each frequency there
are two kink modes, corresponding to the two valleys at $K$ and $K^{\prime}$, for both the
fundamental and the second-harmonic components. Although one can excite the fundamental wave
unidirectionally exploiting the inherent chirality of the kink modes, e.g., by using sources of
either right- or left-circularly polarized light, generally the generated second-harmonic waves
will have both forward- and backward-propagating components.

The intrinsic, local chirality of the kink modes can be characterized by Stokes parameters defined
for the magnetic field $\mathbf{H}=(H_x,H_y)$ as $D=S_3/S_0$, with $S_0=|H_x|^2+|H_y|^2$ and
$S_3=-2\textrm{Im}(H_xH_y^*)$ (for details, see \cite{supp_mat}). The chirality maps for both the
fundamental and harmonic waves at the marked points of $\bar{k}_f$ and $\bar{k}_s^+$ in
Fig.~\ref{fig:fig4}(a) are shown in the right panels of this figure. Guided by this map, one can
see that kink modes can be excited unidirectionally, with an example being given in
Fig.~\ref{fig:fig4}(b), where the three kink modes participating in the SHG process and the
frequency are indicated in Fig.~\ref{fig:fig4}(a) by dots and the blue line, respectively (see
\cite{supp_mat} for further details). While the fundamental wave at $\bar{k}_f$ is launched
unidirectionally rightwards (the source is marked by the arrow) \cite{wyl20ol,ylb20jstqe}, from the
field of  $|E_2|$, one can see that indeed both forward- and backward-propagating
waves are generated.

To analyze the SHG process quantitatively, we perform the Fourier transform of the fields $E_1$ and $E_2$ in
Fig.~\ref{fig:fig4}(b) and present the results in Figs.~\ref{fig:fig4}(c) and \ref{fig:fig4}(d). In
particular, the peak of $|\tilde{E}_1^k|$ corresponds to $\bar{k}_f$ as the fundamental wave is
excited unidirectionally. Interestingly, we can see three peaks in the spectrum of
$|\tilde{E}_2^k|$, two of which correspond to $\bar{k}_s^-$ and $\bar{k}_s^+$, i.e., the forward-
and backward-propagating waves at the second harmonic are generated due to their (near
phase-matched) nonlinear interaction with the fundamental wave at $\bar{k}_f$. The peak at
$2\bar{k}_f-2$, on the other hand, corresponds to a \textit{nonlinear umklapp process}. It is due
solely to the nonlinear polarization $P_{2}\sim\chi^{(2)}E_{1}^{2}$, and since it is not a
phase-matched process, the corresponding SHG does not grow exponentially. From the amplitudes of
the peaks located at $\bar{k}_s^-$ and $\bar{k}_s^+$, we can observe that the forward component
with $\bar{k}_s^-$ is much larger, due to a smaller wavevector mismatch; the corresponding
forward-to-backward ratio, $\eta_{\rightleftarrows}=\eta_{\rightarrow}/\eta_{\leftarrow}$, of the
generated second-harmonic waves as a function of frequency is show in Fig.~\ref{fig:fig4}(e). This
analysis also reveals a key feature of the SHG process, i.e. that $\eta_{\rightleftarrows}$ can be
varied simply by tuning the source frequency, a functionality that non-topological nonlinear optics
does not provide. We have also verified that the period $\lambda_p$ in Fig.~\ref{fig:fig4}(b) does
relate to $\bar{k}_f$ and $\bar{k}_s^-$ via $\lambda_p=2\pi/|\Delta \bar{k}|$ with $\Delta
\bar{k}=(2\bar{k}_f-2)-\bar{k}_s^-$, meaning that the observed oscillations are due to the beating
between the two rightward generated waves.

To further confirm that the waves participating in the SHG process corresponding to the full-wave
simulations presented in Fig.~\ref{fig:fig4}(b) are indeed those suggested by the eigenmode
calculations, we scan the excitation frequency [blue line in Fig.~\ref{fig:fig4}(a)] from the
bottom of the frequency matching window to its top and present in Fig.~\ref{fig:fig4}(f) the
extracted kink mode dispersions together with those shown in Fig.~\ref{fig:fig4}(a). The excellent
agreement demonstrates that the SHG observed in the full-wave simulations is indeed due to the
nonlinear coupling of the kink modes located in the two valley gaps. Figure~\ref{fig:fig4}(g)
further shows the extracted wavevector mismatch $\Delta \bar{k}$ from the peaks of
$|\tilde{E}_2^k|$ compared to that obtained from the eigenmode dispersion curves in
Fig.~\ref{fig:fig4}(a). One can again see a good agreement considering possible finite-size effects
and different algorithms used in the full-wave simulation and the eigenmode calculation. Last but
not least, we would also like to note that the valley-Hall kink modes are robust against structural
disorder \cite{Dong_epl2020_disorder}.
\begin{figure}
\includegraphics[width=\columnwidth]{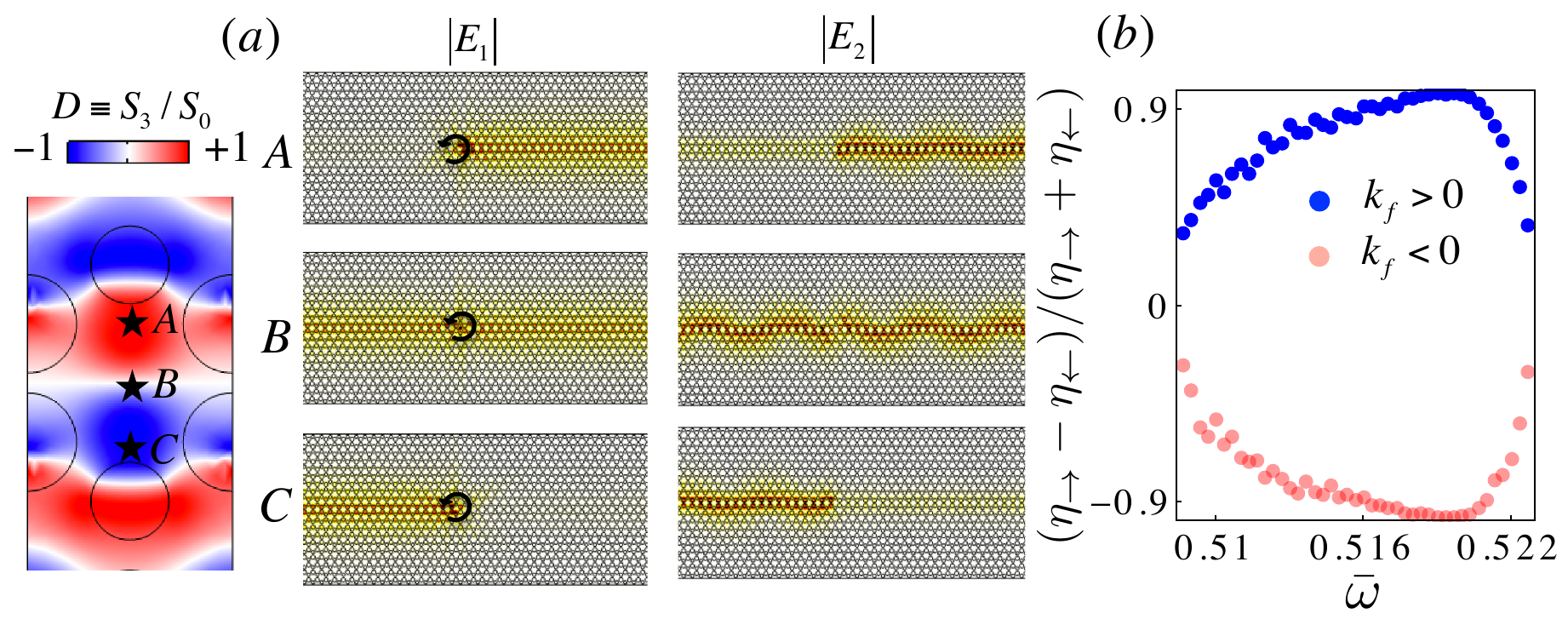}
\caption{\textit{Direction-tunable SHG and nonlinear directional dichroism.} (a) The propagation
direction of the SHG could be switched from rightwards to leftwards by changing the source location
from $A$ to $C$ as labeled in the chirality map. (b) The SHG-DD, defined as
$(\eta_{\rightarrow}-\eta_{\leftarrow})/(\eta_{\rightarrow}+\eta_{\leftarrow})$, determined for
$k_f>0$ ($k_f<0$), corresponding to the case $A$ ($C$) in (a). \label{fig:fig5}}
\end{figure}

\textit{Application to enhanced enantioselectivity in probing chiral molecules.---} Our system
exhibits several new features that could lead to active devices with new functionalities. For
example, the chirality map suggests that one could tune the direction of the SHG by simply changing
the source location of the fundamental wave (location of chiral molecules) as demonstrated in
Fig.~\ref{fig:fig5}(a). Importantly, the plots in Fig.~\ref{fig:fig5}(b) show that using our
proposed device one can achieve remarkably large \textit{SHG directional dichroism} (SHG-DD),
defined as $(\eta_{\rightarrow}-\eta_{\leftarrow})/(\eta_{\rightarrow}+\eta_{\leftarrow})$. In
particular, the maximum value of the SHG-DD, achieved for $\bar{\omega}=0.52$, is as large as 0.97.
This is much larger than what can be achieved with optical systems employing superchiral light
\cite{tc10prl,tc11s,sdh12prx} or comparatively much bulkier nonlinear metasurfaces
\cite{vbc14am,czw16am,kyh20nl} (see \cite{supp_mat} for a detailed quantitative characterization of
such optical probes for sensing chiral molecules).

\textit{Conclusion and outlook.---} In conclusion, we have demonstrated tunable bi-directional SHG
via nonlinear interaction of topological valley-Hall kink modes within two valley gaps in
all-dielectric PhC structures. The ideas presented here could be extended to other nonlinear
optical processes, e.g., third-harmonic generation or four-wave mixing. Implementing nonlinear
frequency mixing processes using quantum spin Hall edge modes \cite{QSH_15PRL_Wu} could also lead to new physics. Moreover, one
could explore other platforms, such as waveguide arrays \cite{valleyWG_PRL18} and coupled
resonators \cite{Loop_NP13_Hafezi}, to achieve similar nonlinear
topological physics to these observed in the current setup.

\textit{Acknowledgments.---} This work was supported by the European Research Council (ERC) (Grant
no. ERC-2014-CoG-648328). W. E. I. S. acknowledges support from Marie Sklodowska-Curie Individual
Fellowship (MSCA-IFEF-ST-752898).


\begin{thebibliography}{99}

\bibitem{review_NP14}L. Lu, J. D. Joannopoulos, and M. Soljacic, Nat. Photonics {\bf 8}, 821 (2014).
\bibitem{review_NP17}A. B. Khanikaev and G. Shvets, Nat. Photonics {\bf 11}, 763 (2017).
\bibitem{review_PQE17}X.-C. Suna, C. He, X.-P. Liu, M.-H. Lu, S.-N. Zhu, and Y.-F. Chen, Progress in Quantum Electronics {\bf 55}, 52 (2017).
\bibitem{review_AOM17} Y. Wu, C. Li, X. Hu, Y. Ao, Y. Zhao, and Q. Gong, Adv. Optical Mater. {\bf 5}, 1700357 (2017).
\bibitem{review_OE18} B.-Y. Xie, H.-F. Wang, X.-Y. Zhu, M.-H. Lu, Z. D. Wang, and Y.-F. Chen, Opt. Express {\bf 26}, 24531 (2018).
\bibitem{review_JAP19} M. S. Rider, S. J. Palmer, S. R. Pocock, X. Xiao, P. A. Huidobro, and V. Giannini, J. Appl. Phys. {\bf 125}, 120901 (2019).
\bibitem{review_RMP19}T. Ozawa, H. M. Price, A. Amo, N. Goldman, M. Hafezi, L. Lu, M. C. Rechtsman, D. Schuster, J. Simon, O. Zilberberg, and I. Carusotto, Rev. Mod. Phys. {\bf 91}, 015006 (2019).

\bibitem{QHE_PRL08_Haldane}F. D. M. Haldane and S. Raghu, Phys. Rev. Lett. {\bf 100}, 013904 (2008).
\bibitem{QHE_PRL08_Wang} Z. Wang, Y. D. Chong, J. D. Joannopoulos, and M. Soljacic, Phys. Rev. Lett. {\bf 100}, 013905 (2008).
\bibitem{QHE_Nature09_Wang} Z. Wang, Y. D. Chong, J. D. Joannopoulos, and M. Soljacic, Nature {\bf 461}, 772 (2009).
\bibitem{QHE_PRL11_Poo} Y. Poo, R.-X. Wu, Z. Lin, Y. Yang, and C. T. Chan, Phys. Rev. Lett. {\bf 106}, 093903 (2011).

\bibitem{Floquet_Nature13_Rechtsman} M. C. Rechtsman, J. M. Zeuner, Y. Plotnik, Y. Lumer, D. Podolsky, F. Dreisow, S. Nolte, M. Segev, and A. Szameit, Nature {\bf 496}, 196 (2013).
\bibitem{Loop_NP13_Hafezi}M. Hafezi, S. Mittal, J. Fan, A. Migdall, and J. M. Taylor, Nat. Photon. {\bf 7}, 1001 (2013).

\bibitem{QSH_NM13_Khanikaev} A. B. Khanikaev, S. H. Mousavi, W. K. Tse, M. Kargarian, A. H. MacDonald, and G. Shvets, Nat. Mater. {\bf 12}, 233 (2013).
\bibitem{QSH_PRL15_Ma}T. Ma, A. B. Khanikaev, S. H. Mousavi, and G. Shvets, Phys. Rev. Lett. {\bf 114}, 127401 (2015).
\bibitem{QSH_PNAS16_He} C. He, X.-C. Sun, X.-P. Liu, M.-H. Lu, Y. Chen, L. Feng, and Y.-F. Chen, Proc. Natl. Acad. Sci. U.S.A {\bf 113}, 4924 (2016).
\bibitem{QSH_15PRL_Wu} L. H. Wu and X. Hu, Phys. Rev. Lett. {\bf 114}, 223901 (2015).
\bibitem{QVH_NJP16_Ma} T. Ma and G. Shvets, New J. Phys. {\bf 18}, 025012 (2016).

\bibitem{QVH_NC19_He}X. T. He, E. T. Liang, J. J. Yuan, H. Y. Qiu, X. D. Chen, F. L. Zhao, and J. W. Dong, Nat. Commun. {\bf10}, 872 (2019).
\bibitem{QVH_NN19_Shalaev} M. I. Shalaev, W. Walasik, A. Tsukernik, Y. Xu, and N. M. Litchinitser, Nat. Nanotechnol. {\bf 14}, 31 (2019).
\bibitem{QVH_LPR19_Ma} J. Ma, X. Xi, and X. Sun, Laser Photonics Rev. {\bf 13}, 1900087 (2019).



\bibitem{review_RPP18_Rachel}S. Rachel, Rep. Prog. Phys. {\bf 81}, 116501 (2018).
\bibitem{book_Boyd08}R. W. Boyd, Nonlinear Optics (Academic Press; 3 edition 2008).
\bibitem{review_NTP} D. Smirnova, D. Leykam, Y. Chong, and Y. Kivshar, Appl. Phys. Rev. {\bf 7}, 021306 (2020).

\bibitem{solition_PRL13_Lumer}Y. Lumer, Y. Plotnik, M. C. Rechtsman, and M. Segev, Phys. Rev. Lett. {\bf 111}, 243905 (2013).
\bibitem{solition_PRL16_Leykam}D. Leykam and Y. D. Chong, Phys. Rev. Lett. {\bf 117}, 143901 (2016).
\bibitem{solition_Mukherjee}S. Mukherjee, and M. C Rechtsman, Science {\bf 368}, 856 (2020).
\bibitem{control_PRL19} D. A. Dobrykh, A. V. Yulin, A. P. Slobozhanyuk, A. N. Poddubny, and Y. S. Kivshar, Phys. Rev. Lett. {\bf 121}, 163901 (2018).
\bibitem{imaging_PRL19} D. Smirnova, S. Kruk, D. Leykam, E. Melik-Gaykazyan, D.-Y. Choi, and Y. Kivshar, Phys. Rev. Lett. {\bf 123}, 103901 (2019).

\bibitem{amplifier_PRX16}V. Peano, M. Houde, F. Marquardt, and A. A. Clerk, Phys. Rev. X {\bf 6}, 041026 (2016).

\bibitem{TIlaser_Science18T}G. Harari, M. A. Bandres, Y. Lumer, M. C. Rechtsman, Y. D. Chong, M. Khajavikhan, D. N. Christodoulides, M. Segev, Science {\bf 359}, eaar4003 (2018).
\bibitem{TIlaser_Science18E}M. A. Bandres, S. Wittek, G. Harari, M. Parto, J. Ren, M. Segev, D. N. Christodoulides, and M. Khajavikhan, Science {\bf 359}, eaar4005 (2018).

\bibitem{quantum_Nature18}S. Mittal, E. A. Goldschmidt and M. Hafezi, Nature {\bf 561}, 502 (2018).


\bibitem{harmonic_OE18} C. Qian, K. H. Choi, R. P. H. Wu, Y. Zhang, K. Guo, and K. H. Fung, Optics Express {\bf 26}, 5083 (2018).
\bibitem{harmonic_NC19}Y. Wang, L.-J. Lang, C. H. Lee, B. Zhang and Y. D. Chong, Nat. Commun. {\bf 10}, 1102 (2019).
\bibitem{harmonic_NN19}S. Kruk, A. Poddubny, D. Smirnova, L. Wang, A. Slobozhanyuk, A. Shorokhov, I. Kravchenko, B. L.-Davies, and Y. S. Kivshar, Nat. Nanotechnol. {\bf 14}, 126 (2019).

\bibitem{SciAdv20_You} J. W. You, Z. Lan, and N. C. Panoiu, Sci. Adv. {\bf 6}, eaaz3910 (2020).
\bibitem{PRB20_Lan} Z. Lan, J. W. You, and N. C. Panoiu, Phys. Rev. B {\bf 101}, 155422 (2020).


\bibitem{HoneyValley_PRB17dong}X.-D. Chen, F.-L. Zhao, M. Chen, and J.-W. Dong, Phys. Rev. B {\bf 96}, 020202(R) (2017).
\bibitem{HoneyValley_SciRe18Hang}Y. Yang, H. Jiang and Z. H. Hang, Sci. Rep. {\bf 8}, 588 (2018).

\bibitem{RSoft}BandSOLVE; www.synopsys.com.

\bibitem{Chern_ZhaoOE20} R. Zhao, G.-D. Xie,  M. L. N. Chen, Z. Lan, Z. Huang, and W. E. I. Sha, Opt. Express {\bf 28}, 4638 (2020).
\bibitem{Chern_LuFO20}C. Wang, H. Zhang, H. Yuan, J. Zhong, and C. Lu, Front. Optoelectron. {\bf 13}, 73 (2020).
\bibitem{Chern_PazAQT20}M. B. de Paz, C. Devescovi, G. Giedke, J. J. Saenz, M. G. Vergniory, B. Bradlyn, D. Bercioux and A. G. Etxarri, Adv. Quantum Technol. {\bf 3}, 1900117 (2020).
\bibitem {supp_mat} See Supplemental Material at XXX.
\bibitem{wyl20ol}Y. Wang, J. W. You, Z. Lan, and N. C. Panoiu, Opt. Lett. \textbf{45}, 3151 (2020).
\bibitem{ylb20jstqe}J. W. You, Z. Lan, Q. Bao, and N. C. Panoiu, IEEE J. Sel. Top. Quantum Electron. \textbf{26}, 4600308 (2020).

\bibitem{Dong_epl2020_disorder} Z. Dong, F. Xu, and W. Liang, Europhys. Lett. {\bf 131}, 54002 (2020).

\bibitem{tc10prl}Y. Tang and A. E. Cohen, \prl \textbf{104}, 163901 (2010).
\bibitem{tc11s}Y. Tang and A. E. Cohen, Science \textbf{332}, 333 (2011).
\bibitem{sdh12prx} M. Schaferling, D. Dregely, M. Hentschel and H. Giessen, Phys. Rev. X \textbf{2}, 031010
(2012).

\bibitem{vbc14am}V. K. Valev, J. J. Baumberg, B. De Clercq, N. Braz, X. Zheng, E. J. Osley, S.
Vandendriessche, M. Hojeij, C. Blejean, J. Mertens, C. G. Biris, V. Volskiy, M. Ameloot, Y. Ekinci,
G. A. E. Vandenbosch, P. A. Warburton, V. V. Moshchalkov, N. C. Panoiu, and T. Verbiest, Adv.
Mater. \textbf{26}, 4074 (2014).
\bibitem{czw16am}S. Chen, F. Zeuner, M. Weismann, B. Reineke, G. Li, V. K. Valev, K. W. Cheah, N.
C. Panoiu, T. Zentgraf, and S. Zhang, Adv. Mater. \textbf{28}, 2992 (2016).
\bibitem{kyh20nl}D. Kim, J. Yu, I. Hwang, S. Park, F. Demmerle, G. Boehm, M. C. Amann, M. A. Belkin, and J. Lee, Nano Lett. \textbf{20}, 8032 (2020).

\bibitem{valleyWG_PRL18}J. Noh, S. Huang, K. P. Chen, and M. C. Rechtsman, Phys. Rev. Lett. {\bf 120}, 063902 (2018).


\end{thebibliography}
\end{document}